\documentclass[usenatbib]{mn2e}
\input{epsf}
\usepackage{amsmath,amssymb}

\def\lsim{ \lower .75ex \hbox{$\sim$} \llap{\raise .27ex \hbox{$<$}} }
\def\gsim{ \lower .75ex \hbox{$\sim$} \llap{\raise .27ex \hbox{$>$}} }
\newcommand{\apjl}{ApJ}
\newcommand{\apj}{ApJ}
\newcommand{\araa}{ARA\&A}

\newcommand{\mnras}{MNRAS}

\newcommand{\aap}{A\&A}

\newcommand{\pasj}{PASJ}
\newcommand{\nat}{Nature}
\newcommand{\prd}{Phys.Rev.D}

\begin{document}
 
\title[EROs clustering in a hierarchical universe]
{Massive, red galaxies in a hierarchical universe-II 
Clustering of Extremely Red Objects}

\author[Gonzalez-Perez  et al.]{
\parbox[t]{\textwidth}{
\vspace{-1.0cm}
V.\,Gonzalez-Perez$^{1}$,
C.\,M.\,Baugh$^{1}$,
C.\,G.\,Lacey$^{1}$,
J.-W.\,Kim$^{1}$.
}
\\
$^{1}$Institute for Computational Cosmology, Department of Physics, 
University of Durham, South Road, Durham, DH1 3LE, UK.
\\
}
 
\maketitle
\begin{abstract}
We present predictions for the clustering of Extremely Red Objects 
(EROs) in a $\Lambda$ cold dark matter Universe, using a
semi-analytical galaxy formation model in combination with a cosmological N-body simulation. EROs are red, massive galaxies 
observed at $0.7\lesssim z\lesssim 3$, and their numbers and properties
have posed a challenge to hierarchical galaxy formation models. We analyse the halo occupation
distribution and two-point correlation function of EROs, exploring how
these quantities change with apparent magnitude, colour cut and
redshift. Our model predicts a halo occupation distribution that is
significantly different from that typically assumed. This is due to
the inclusion of AGN feedback, which changes the slope and scatter of
the luminosity-host halo mass relation above the mass where AGN
feedback first becomes important.
We predict that, on average, dark matter haloes with masses above
$10^{13}h^{-1}M_{\odot}$ host at least one ERO at $1.5\le z \le 2.5$. Taking
into account sample variance in observational estimates, the predicted angular clustering for EROs with
either $(R-K)>5$ or $(i-K)>4.5$ is in reasonable agreement with
current observations. 
\end{abstract}

\section{Introduction}
\defcitealias{eros1}{Paper I}
%
%
%

Hierarchical models of galaxy formation have enjoyed many 
successes at both low and high redshifts. However, such 
models are still challenged by the large number of galaxies 
observed with old stellar populations, many of which were already 
in place at redshifts higher than $z=1$ when the Universe was 
less than half its present age \citep[e.g.][]{k201,vaisanen04}. 
Observationally it is possible to select such galaxies by requiring
them to have very red optical and near-infrared colours. Extremely Red Objects
(EROs) are commonly classified as galaxies with colours redder than
$(R-K)_{\rm Vega}=5$ \citep[see e.g.][]{mccarthy04}. This
classification was originally thought to select galaxies at very 
high redshift \citep[$z>5$,][]{elston88}. However, later 
spectroscopic confirmation showed that EROs were actually at 
$0.8\lesssim z\lesssim 2$ \citep{cimatti02b,cimatti02a,conselice07,wilson07}.

Currently there are no direct measurements of the spatial two-point correlation function 
of EROs, due to the small size of available surveys. The constraints 
on the clustering strength of EROs come from deprojecting angular clustering. Assumptions are required 
to extract the spatial correlation length from such measurements. 
Nevertheless, a consensus has been reached in that the observed 
clustering of EROs appears to be strong, with inferred correlation 
lengths comparable to those of early type galaxies 
today \citep{daddi00xi,daddi01,daddi02,firth02,
roche02,miyazaki03,brown05,kong06,kong09,kim11}. This implies that 
these objects are more strongly clustered than the underlying dark 
matter at the typical median redshift of these samples, which is close
to $z = 1$. Improved measurements of ERO clustering require 
larger areas of the sky to be mapped. This is being achieved by 
various near-infrared surveys that are pushing ERO clustering 
measurements out towards larger angular pair separations \citep{kim11}.

Measurements of the clustering strength of EROs and its dependence 
on properties such as luminosity and colour will allow us 
to pin down the masses of the dark matter haloes which host these 
galaxies and hence to constrain the physics behind their formation. 
The high clustering amplitude for EROs implied by current observations 
suggests that these galaxies reside in massive haloes, in agreement 
with the predictions in the first paper in this series 
\citep[][hereafter Paper I]{eros1}. The clustering signal also provides 
new constraints on the form of the feedback from both supernovae and
active galactic nuclei (AGN) beyond those provided by the abundance of
EROs.

The clustering of massive red galaxies has previously been interpreted 
using empirical models. \citet{moustakas02} used the halo mass 
function and a power law form for the halo occupation distribution (HOD) 
to generate a simple model tuned to reproduce observations of EROs 
with $H<20.5$ and $(I-H)>3$ by \citet{firth02}. 
\citet{tinker10} also used a halo occupation distribution model 
to interpret the clustering of distant red galaxies (DRG), which typically 
have a higher redshift ($z \sim 2$) than EROs. Such models allow one
to estimate the halo mass in which red galaxies are found, without addressing the 
physics behind how the galaxies formed or why they have red colours. 
These calculations depend on assumptions about the form of the 
halo occupation distribution. As we shall show, this is a difficult
proposition as the HOD of EROs 
is different from that of less extreme, better studied galaxies
\citep[see also][]{almeida10}.  

The abundance of EROs has also been tackled in more ambitious 
calculations which do attempt to address the origin of EROs and 
their photometric properties \citep{nagamine05,kang06,fontanot10}. 
\citet{nagamine05} used mesh and particle based gas dynamic 
simulations to study massive galaxies at the redshifts of EROs. The 
simulation boxes used were relatively small, as demanded by the 
need to attain reasonable spatial and mass resolution in the 
calculations. This makes it difficult to study rare populations 
with strong clustering using this approach. In order to match 
the observed abundance of EROs, \citeauthor{nagamine05} were forced to apply, 
by hand, quite a high level of extinction to all of their galaxies. 
Thus, all of their EROs are dusty star forming galaxies by construction, 
with no information about the number of galaxies with red colours 
due to their having old stellar populations.  
Semi-analytical galaxy formation models have also been applied to 
study EROs. A key consideration here is that the model should reproduce 
the properties of galaxies in the local universe, which are used to 
set most or all of the model parameters \citep{baugh06}. Of particular relevance for 
ERO predictions is the present day K-band luminosity function. The 
bright end of the luminosity function is dominated by early-type galaxies \citep{norberg02}. The challenge is for a model to match the abundance of 
EROs at intermediate redshift whilst still producing the observed number of 
bright galaxies locally. Both the \citet{kang06} and \citet{fontanot10} 
models overpredict the number of bright galaxies in the K-band today. In
particular, the \citet{fontanot10} model overpredicts the number of
galaxies by a factor of $10$ at $M_{K}-5{\rm log}\,h=-25.8$, which corresponds to the knee 
of the luminosity function. 

In this paper we make predictions for the clustering of EROs based on the published semi-analytical 
model of \citet{bower06} in its original form. This model reproduces
many observations, including the present-day K-band luminosity 
function and the number counts of galaxies with $15\leq K\leq 25$, and 
the inferred evolution of the stellar mass function. 
In \citetalias{eros1} we showed that the \citeauthor{bower06} model 
successfully reproduces the observed number counts of EROs,
without having to adjust any of its parameters. The dust extinction 
is calculated self-consistently rather than being put in by hand 
(see Cole et~al. 2000). 
We also predicted that EROs should be the most massive and brightest of 
K-selected galaxies at $1\lesssim z\lesssim 2$, with a population 
dominated by old, passively evolving galaxies. Our predictions 
broadly agree with observations of EROs. We found that AGN
feedback is a key ingredient for understanding the evolution of the
most massive galaxies which were in place at $z\sim 1$. \citeauthor{fontanot10} note that the
modelling of stellar populations also plays a significant role in shaping the
predictions for EROs \citep[see also][]{tonini09}.

In this paper we study the halo occupation distribution and clustering
of EROs. Previously, \citet{qiguo09} reproduced the clustering of a
similarly selected galaxy population, DRGs, using a different semi-analytical model. \citeauthor{qiguo09} 
restricted their study to a single magnitude limited 
sample of galaxies. Here we will explore the variation of ERO 
clustering with colour, magnitude, redshift, mode of star formation and
also investigate how the clustering changes from real to redshift space, trying to
understand the physical processes behind each change.

In this paper we select EROs using either their $(R-K)$ or $(i-K)$ colours. 
Colour cuts using $(i-K)$ and also $(I-K)$ appear to
be more effective than $(R-K)$ colours at removing ``low'' redshift 
galaxies with $z<0.8$ and isolating galaxies at higher redshifts
\citep{conselice08,kong09}. \citet{kim11} found that the
clustering of EROs selected by their $(R-K)$ or $(i-K)$ colours
is different, particularly at larger pair separations, where
$(i-K)$ selected galaxies have larger clustering amplitudes.

The paper is organised as follows. In Section
\ref{sec:model}, we summarise the key features of
\citeauthor{bower06} galaxy formation model. The predictions for the
halo occupation distribution of EROs are presented in Section
\ref{sec:hod}. The predictions for the spatial clustering are
discussed in Section \ref{sec:xi} and  those for the angular
clustering in Section \ref{sec:w}. Conclusions can be found in Section \ref{sec:conclusions}.

\section{Galaxy formation model}\label{sec:model}

We predict the clustering of EROs in a $\Lambda$CDM 
universe using the {\sc galform} semi-analytical galaxy formation 
model developed by \citet{cole00}. Semi-analytical models 
use simple, physically motivated recipes and rules to follow the 
fate of baryons in a universe in which structure grows hierarchically 
through gravitational instability \citep[see][for an overview of 
hierarchical galaxy formation models]{baugh06}. 

{\sc galform} follows the main processes which shape the formation and
evolution of galaxies. These include: (i) the collapse and merging of
dark matter haloes; (ii) the shock-heating and radiative cooling of
gas inside dark matter haloes, leading to the formation of galaxy
discs; (iii) quiescent star formation in galaxy discs; (iv) feedback
from supernovae, from active
galactic nuclei (AGN) and from photoionization of the intergalactic medium (IGM); (v) chemical enrichment of the stars and gas; (vi) galaxy
mergers driven by dynamical friction within common dark matter haloes,
leading to the formation of stellar spheroids, which also may trigger
bursts of star formation. The end product of the calculations is a
prediction for the number and properties of galaxies that reside
within dark matter haloes of different masses.

In this paper we focus our attention on the \citet{bower06}
variant of {\sc galform}. In \citetalias{eros1} we showed that this model reproduces the observed
numbers of EROs, whereas the \citet{baugh05} model underpredicted the
counts of these galaxies. The \citeauthor{bower06} model is implemented in the Millennium Simulation
\citep{springel05}: an N-body simulation with about
$10^{10}$ particles, each with a mass of $8.6\times
10^{8}h^{-1}$M$_{\odot}$, in a box of side $500h^{-1}$Mpc. The galaxy
population is output at selected snapshots in the Millennium
Simulation. The outputs are evenly spread in the logarithm of the
expansion factor and so do not correspond to round numbers in
redshift, as will become apparent when we plot our predictions
(e.g. Fig. \ref{fig:usahod} shows predictions at $z=2.1$ rather than $z=2$).

Key features of the \citeauthor{bower06} model include (i) a time scale for quiescent
star formation that scales with the dynamical time of the disk and which
therefore changes significantly with redshift \citep[see][for a study of
  different star formation laws in quiescent galaxies]{lagos10}, (ii) bursts of star formation occur due
to both galaxy mergers and when disks become dynamically unstable, and (iii) feedback
from both supernovae and AGN \citep[see][for a discussion of
  the effect that feedback has on the luminosity function of
  galaxies]{benson03}. The onset of the AGN suppression of the cooling
flow is governed by a comparison of the cooling time of the gas with
the free-fall time for the gas to reach the centre of the
halo. Cooling is suppressed in quasi-static hot haloes if the
luminosity released by material accreted onto a central supermassive
black hole balances the cooling luminosity \citep[see][for a full
  description of black hole growth in the model]{nikos11}. \citeauthor{bower06} adopt the 
cosmological parameters of the Millennium Simulation \citep{springel05}, 
which are in broad agreement with constraints from measurements 
of the cosmic microwave background radiation and large scale galaxy clustering 
\citep[e.g.][]{sanchez09}: $\Omega_{0}=0.25$, $\Lambda_{0} = 0.75$, 
$\Omega_{b}=0.045$, $\sigma_{8}=0.9$ and $h=0.73$, such that the Hubble constant
today is $H_0=100\,h$ km$\,{\rm s}^{-1}$Mpc$^{-1}$.  The
\citeauthor{bower06} model parameters 
were fixed with reference to a subset of the available observations 
of galaxies, mostly at low redshift \citep[see][for a discussion of
  parameter fitting]{bower10}. This model successfully
reproduces the inferred stellar mass function up to $z=4.5$. For further details
we refer the reader both to \citetalias{eros1} in this series and to \citet{bower06}. 

Here we extract predictions for the clustering of EROs using the \citeauthor{bower06} model, without 
adjusting any of its parameters.

The bands used here correspond to the R band from SUBARU, the i band
from the SDSS and the K band from UKIRT, 
with effective wavelengths of $0.65$ $\mu$m, $0.75$ $\mu$m and $2.2$ $\mu$m, respectively.  All magnitudes used in this
paper are in the Vega system, unless otherwise specified. The Vega-AB
offsets are $-0.196,\, -0.355$ and $-1.87$ magnitudes for the R, i and K bands, respectively.

\section{The halo occupation distribution of EROs and the abundance of
their host haloes}\label{sec:hod}

In this section we explore first the halo occupation distribution
(HOD) of EROs and then their HOD-weighted halo mass function, which
helps to relate the HOD to the clustering predictions.

\subsection{The HOD of EROs}
The spatial clustering of galaxies can be quantified by the
correlation function, i.e. the excess or deficit of the number of
galaxy pairs at a given separation with respect to a
random distribution. The correlation function of galaxies can be
derived from the halo occupation
distribution once the mass function and clustering of the haloes are
known \citep[e.g.][]{benson00}. The steps
relating the HOD to the effective bias of a galaxy sample have been
reviewed by \citet{han09}. 

The HOD gives, as a
function of halo mass, the mean
number of galaxies per halo, $\langle N \rangle_{M}$, which pass a particular
observational selection. Semi-analytical models
naturally predict the form of the HOD, which is determined by the interplay between a
range of physical processes which can change with redshift. The form of the HOD can be understood in terms of the sum of
  the contribution of {\it central}\footnote{In hierarchical models the {\it central} galaxy is defined
  as the most massive one within a dark matter halo. This galaxy is, generally, situated
  close to the centre of mass of the halo \citep{benson00b,benson00,berlind03}.
} galaxies, which typically is assumed to resemble a step function, and that of {\it satellite} galaxies,
which approximately follows a power law \citep{berlind03,zheng05}. This implies that,
  in general, $\langle N \rangle_{M}$ has a cutoff at low masses and then a slow
  rise or plateau followed by a power law with increasing halo mass
  \citep{benson00b,benson00,berlind02}. However, the HOD plateau has
  been found to be less evident for brighter galaxies \citep{zehavi10}
  and the same is expected to happen for an older
population of galaxies \citep[][note, however, that these theoretical
  studies were developed before the inclusion of AGN feedback in galaxy formation models]{berlind03,zheng05}.

\begin{figure}
	\begin{center}
	{\epsfxsize=8.5truecm
	\epsfbox[56 12 545 473]{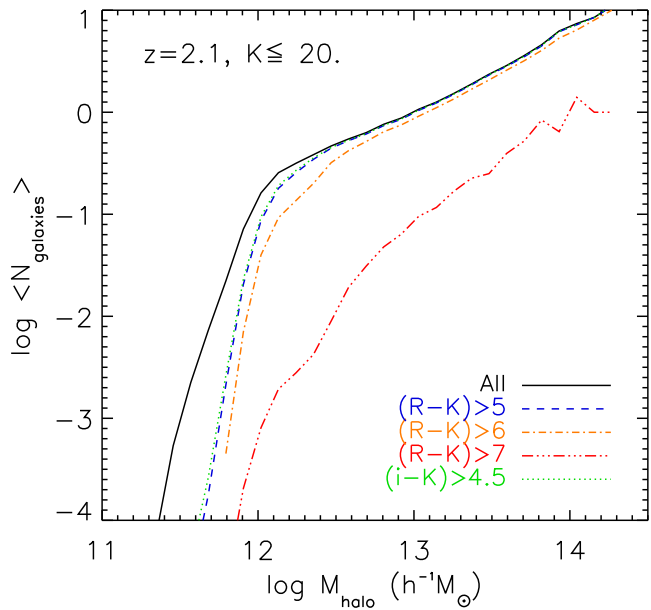}}
	\caption{The HOD at $z=2.1$ of galaxies with $K\leq 20$ (solid
          black line) and of EROs with the same K limit and $(R-K)>5,6,7$ (dashed, dash-dotted and dash-triple dotted
          lines lines respectively, as indicated by the key) and $(i-K)>4.5$ (dotted
          line).  
        }
	\label{fig:usahod}
	\end{center}
\end{figure}

\begin{figure}
	\begin{center}
	{\epsfxsize=8.5truecm
	\epsfbox[33 26 525 462]{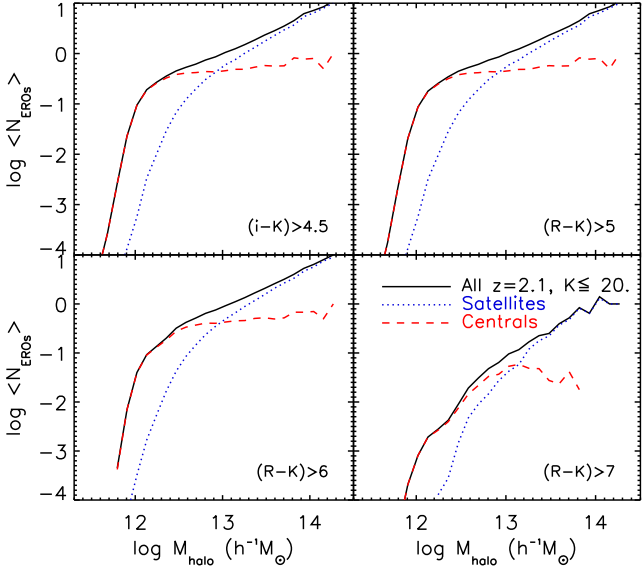}}
	\caption{Solid lines show the HOD of all EROs with $K\leq 20$
          at $z=2.1$. The dotted (dashed) lines show the
          predicted HOD for satellite (central) EROs. Different panels
          show the HOD predictions for EROs selected with different colour cuts, as labelled.
        }
	\label{fig:sat_usahod}
	\end{center}
\end{figure}


In Fig. \ref{fig:usahod} the predicted HOD of EROs with $K\leq 20$ at $z=2.1$ is
compared with that of all
galaxies with the same magnitude limit, but without any colour cut. At
this redshift, the
HOD of K-selected galaxies and of EROs with either
$(R-K)>5$ or $(i-K)>4.5$ differ only in the low mass cutoff. As we found in
\citetalias{eros1}, this shows that EROs with $K \leq 20$ tend not to be
hosted by haloes with low masses ($M_{\rm halo}>4\times
10^{11}h^{-1}M_{\odot}$ in this case). At $z=2.1$, the cut in apparent
magnitude of $K=20$, corresponds to an absolute magnitude of
$M_{K}-5{\rm log}\,h\approx
-25.5$ (observer frame). According to the model, this value is close to $L*$ at $z=2.1$. Thus, Fig. \ref{fig:usahod} shows the
HOD of $\sim L*$ and
brighter galaxies. In \citetalias{eros1} it was shown that in the
\citeauthor{bower06} model, EROs
account for most of the bright end of the K-band luminosity function at
$1\lesssim z\lesssim 2.5$, explaining the similar HODs
predicted for EROs and for all K-selected galaxies at $z=2.1$. At lower
redshifts in the ERO range, moving toward $z=1$, the difference
between the ERO HOD and the overall galaxy HOD becomes larger. In this
case, we are picking up less luminous galaxies and a smaller fraction of
these satisfy the colour cut to be classified as EROs.

Fig. \ref{fig:usahod} shows  that the HOD of all galaxies with
$K \leq 20$ at $z=2.1$ is predicted to have a
clear cutoff at low masses followed by an approximately power law
rise. Below $M_{\rm halo}\sim 10^{12}h^{-1}M_{\odot}$, the probability of hosting
an ERO is close to zero. Nevertheless, as we shall see in the next
subsection, such haloes where the HOD is rising from zero can have a big
influence on the predicted clustering \citep[see][]{han09}. To model this transition region accurately it is necessary to generate many examples of merger histories of
such haloes, to accurately quantify the fraction that host EROs. 

Fig. \ref{fig:usahod} shows  that the HOD of all galaxies with $K \leq
20$ at $z=2.1$ flattens out at a value of $\langle N \rangle _{M}$
below unity, in contradiction with the expectation for the generic form
of the HOD. This implies that
central galaxies with these characteristics are not expected to be
found in every halo. The same trend is found for
EROs. Fig. \ref{fig:sat_usahod} shows that for EROs at $z=2.1$, the HOD of central galaxies only reaches
unity for haloes with $M>10^{14}h^{-1}M_{\odot}$, except in the case
of EROs with $(R-K)>7$, for which a mean occupation of unity is never
reached. 

To understand this prediction, we examine the trend between the K-band luminosity
and the host halo mass. If there was a well defined trend between the luminosity of
a central galaxy and the mass of its host halo, then the central galaxies
residing in haloes above a given mass would typically be brighter than
some limiting magnitude. We say ``typically'' to allow for scatter in the relation between galaxy luminosity and halo
mass. However, in the \citeauthor{bower06} model, we do not find such a correlation. For
low halo masses, those in which AGN suppression of gas cooling is not
effective, a monotonic relation between central galaxy luminosity and
halo mass is predicted, with sufficiently small scatter to make the
trend clear. This relation flattens and the scatter increases
substantially beyond the mass in which gas cooling is first
suppressed by accretion onto the central supermassive black hole. A
central galaxy of a particular luminosity can be found in a broad
range of halo masses, covering several decades in mass. For models including AGN feedback, it is no longer the case that the brightest central
galaxies will be found exclusively in the most massive
haloes. Instead, the tails of the luminosity-halo mass relation play a
critical role. We find that for the brightest galaxies, only a
fraction of haloes of any mass have formation histories which can
produce such an object, hence $\langle N \rangle _{M}\lesssim 1$.

Fig. \ref{fig:usahod} shows that the minimum mass, $M_{1}$, required for a halo to host
two K-selected galaxies is a factor of $\sim 11$ higher
than that required to host only one, which is labelled $M_{min}$. This is also the case
for EROs except for those with $(R-K)>7$. In this case, haloes are
predicted to host at most one ERO with $(R-K)>7$ and thus, $\langle N \rangle _{M}< 2$ for all masses.
The similarity in the factor found for EROs and K-selected galaxies
indicates that our result is not peculiar to the
additional colour selection associated with an ERO sample, but appears
to be a characteristic of bright galaxies. Indeed, we note that a similar ratio $M_{1}/M_{\rm min}\sim 11$ has been suggested
for local galaxies brighter than $1.1L*$, while $M_{1}/M_{\rm min}\sim 17$
for fainter galaxies \citep{zehavi10}.

Fig. \ref{fig:sat_usahod} compares the predicted HOD for EROs selected
with different colour cuts, separating the contributions from central
and satellite galaxies.
We note in Fig. \ref{fig:sat_usahod} that the step in the HOD is less
pronounced for redder central EROs. As an example, we find at $z=2.1$ an average of
$0.1$ EROs with $(R-K)>5$ and $K\leq 20$ per halo with
$10^{12}h^{-1}M_{\odot}$, while fewer than one in a thousand haloes host EROs with
$(R-K)>7$ and $K\leq 20$ at this mass. In \citetalias{eros1}, we found that redder cuts select brighter and older EROs. Thus, on the one hand, the variation of the HOD with colour predicted here
is in agreement with a $\langle N \rangle _{M}$ shape dominated by
bright and old
galaxies \citep{berlind03,zheng05,han09,zehavi10}. On the other hand,
Fig. \ref{fig:sat_usahod} shows that the split between central and
satellite galaxies varies with the colour cut used to define the ERO
sample. The fraction of satellite galaxies increases for the redder cuts. The difference in the number of satellite EROs causes the
shape of the halo occupation distribution to vary from a clear step
function plus a power law for EROs with
$(R-K)>5$ or $6$, to a distribution close to a pure power law for
EROs with $(R-K)>7$. In this case, the HOD for central galaxies does not get
close to unity before it is dominated by the satellite HOD.

\subsection{The HOD-weighted halo mass function}

\begin{figure}
	\begin{center}
	{\epsfxsize=8.5truecm
	\epsfbox[1 38 343 536]{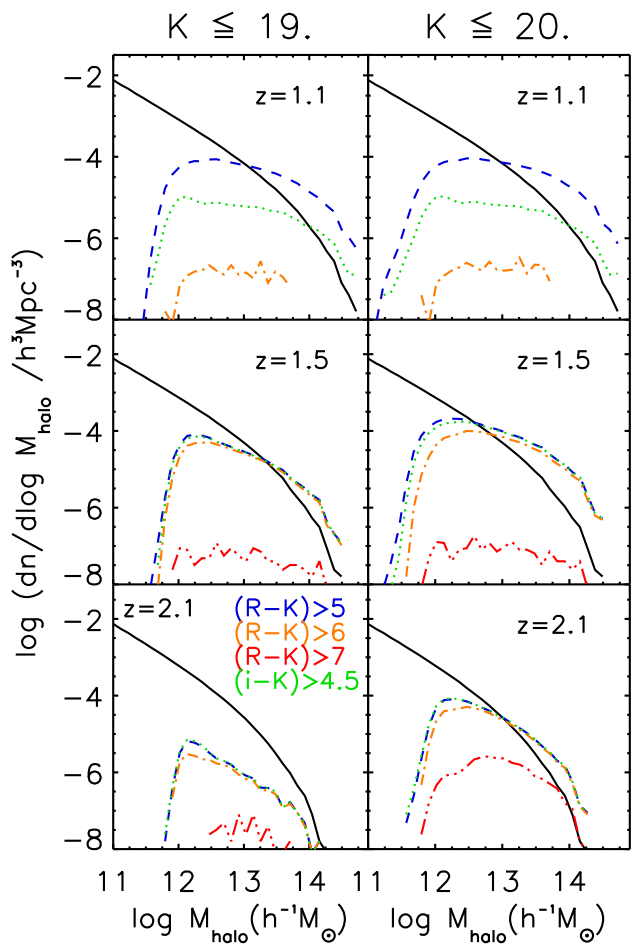}}
	\caption{The mass function of all dark halos (solid lines) and the
          HOD-weighted halo mass function for different ERO selections. As in Fig. \ref{fig:usahod}
          dotted lines correspond to the mass function weighted by EROs
          with $(i-K)>4.5$, dashed lines to $(R-K)>5$,
          dash-dotted to $(R-K)>6$ and dash-triple dotted to
          $(R-K)>7$. Left-hand panels show the prediction for EROs
          with $K\leq 19$ and right-hand ones for $K\leq 20$. Top row
          corresponds to $z=1.1$, the middle to $z=1.5$ and the bottom row
          to $z=2.1$, as labelled. }
	\label{fig:hod}
	\end{center}
\end{figure}

The HOD by itself does not contain any explicit information about the
number density of
dark matter haloes. For a given redshift and assuming an initial power spectrum of matter fluctuations, the number density of
halos as a function of mass, i.e, the halo
mass function, depends only on the cosmology. \citet{mf} argued that
for a Gaussian density field the abundance of
haloes more massive than a characteristic mass is expected to decline
exponentially, with a power law like behaviour at lower masses, a prediction confirmed by numerous N-body simulations
\citep[e.g.][]{jenkins01}.  Fig. \ref{fig:hod} shows as solid black lines the mass function of dark
halos at different redshifts predicted by the Millennium Simulation.

We plot in Fig. \ref{fig:hod} the halo mass function
multiplied by the mean number of EROs per halo, where the EROs are
defined with different colour cuts, $(i-K)>4.5$ and $(R-K)>5,\, 6,\,
7$, and in two magnitude ranges, $K\leq 19$ and $K\leq 20$. As anticipated in \citetalias{eros1},
we see from Fig. \ref{fig:hod} that EROs can be found in
halos with a wide range of masses. In the lowest mass bins we see a sharp
cutoff driven by the shape of the HOD.

We note that a colour cut not shown in Fig. \ref{fig:hod} but that is widely used
for observationally selecting  EROs is $(R-K)>5.3$. At $z=1.1,\, 1.5$
and $2.1$, the HOD for EROs
with  $(i-K)>4.5$ is practically the same as that for EROs
with $(R-K)>5.3$.

Fig. \ref{fig:hod} shows how at $z=1.5$ and $z=2.1$ the weighted mass functions for EROs with
$(i-K)>4.5$ and $(R-K)>5$ practically overlap, while they differ at $z=1.1$. For the two
magnitude limits shown in Fig. \ref{fig:hod}, the predicted redshift
distributions of EROs with either $(i-K)>4.5$ or $(R-K)>5$ have medians
around $z=1.5$. However, the low redshift tail cuts off more sharply for EROs with $(i-K)>4.5$ and, therefore, these EROs are expected to contribute
less to the mass function at $z=1.1$, explaining the difference seen
in Fig. \ref{fig:hod}.

The mass of haloes such that $\langle N \rangle _{M}=1$ for EROs is
given by the mass at
which the HOD-weighted halo mass function has the same amplitude as
the halo mass function. In Fig. \ref{fig:hod} this happens when a line crosses the dark matter
halo mass function (solid line). More massive
halos will contain, on average, more than one ERO. In Fig. \ref{fig:hod} we see that for EROs with $(R-K)>5$ and $K\leq
19$ the mass at which their $\langle N \rangle _{M}=1$ increases by a factor of $\sim 2.5$ from
$z=1.1$ to $z=1.5$. Fig. \ref{fig:hod} shows again that
EROs only inhibit halos more massive than a certain threshold
value, which depends on redshift, apparent magnitude limit and on
colour. Both the mass at which $\langle N \rangle _{M}=1$ and the
minimum mass for a halo to host an ERO increase with redshift. This trend is related to the fixed
apparent magnitude limit which means that we are looking at
intrinsically brighter galaxies at higher redshifts. 


In contrast, the median mass of haloes hosting EROs increases with reducing redshift. For EROs with $(R-K)>5$ and $K\leq 19$ the median mass of their host
haloes increases from $1.8\times 10^{12}\,
h^{-1}M_{\odot}$ at $z=2.1$, to $4.4\times 10^{12}\,
h^{-1}M_{\odot}$ at $z=1.1$. This increase is close to that for the dark matter $M*$. Thus, this tendency is related to the growth of
dark matter haloes with time and it is seen also for EROs with
$(i-K)>4.5$ and for K-selected galaxies. 

We have found that brighter EROs are hosted by more
massive haloes. In particular, at $z=1.5$, halos more massive
than $\sim 2\times 10^{13} h^{-1}M_{\odot}$ will contain, on average,
at least one ERO with $K\leq 19$, while one or more EROs with $K\leq
20$ will be present in halos with masses $\sim 5\times 10^{12}
h^{-1}M_{\odot}$. Our predicted values agree with
those from the empirical model of \citet{moustakas02}.

In general, as shown by Fig. \ref{fig:hod}, the redder an ERO, the
more massive is the typical host halo. In
particular, all halos are predicted to host on average fewer than one ERO
redder than $(R-K)=7$. For EROs with  $K\leq 19$ at $z=1.5$, the median mass of their host
haloes increases from $2.5\times 10^{12}\,
h^{-1}M_{\odot}$ for EROs with $(R-K)>5$, to $6.3\times 10^{12}\,
h^{-1}M_{\odot}$ for EROs with $(R-K)>7$.

We have explored the HOD-weighted mass function separating
the contribution from satellite and central EROs, and find that at
$z=1.1$ central EROs are rare, i.e.  they make a marginal contribution
to the ERO-weighted mass function for massive haloes. The average number of central
EROs at $z=1.1$ is well below one per halo, even for halos as massive as
$10^{14}h^{-1}M_{\odot}$. This is not the case at redshifts $z=1.5$
and $z=2.1$. At these redshifts, we find that massive haloes are expected to host an ERO as their central
galaxy (except at $z=2.1$ and $K\le 19$, in which case EROs are just
rare). Thus, the increase from $z=1.5$ to $z=1.1$ in the halo mass where $\langle N \rangle _{M}=1$ for EROs with $(i-K)>4.5$, appears to be
related to the fact that at lower redshifts the fraction of
satellite EROs increases.

\section{The two-point correlation function of EROs}\label{sec:xi}

In this section we present the predictions for
the two-point correlation function of EROs, $\xi(r)$. The
\citeauthor{bower06} model is implemented in the Millennium
simulation and thus, it can be used to directly predict the spatial distribution of
galaxies and hence the  two-point correlation function. In real-space,
the Cartesian comoving co-ordinates\footnote{
Unless otherwise noted, hereafter we will refer to comoving
co-ordinates simply as coordinates.} 
of EROs within the simulation box are used
to compute pair separations. We calculate their two-point correlation
function, $\xi$, using a simple estimator \citep[e.g.][]{peebles80}: 
\begin{equation}\label{ec:xi}
1+\xi=\frac{2 DD}{n^2V {\rm d} V} \, ,
\end{equation}
where $DD$ is the number of distinct galaxy pairs with
separation between $r$ and $r+{\rm d}r$  measured from the
simulation, allowing for periodic boundary conditions, $n$ is the mean number density of
galaxies, $V$ is the total simulated volume, and d$V$ is the
volume of the spherical shell of radius $r$ and thickness d$r$. The
denominator of Eq. \ref{ec:xi} corresponds to the average number of
neighbours found in d$V$ in the case of a Poisson distribution. 

\subsection{The shape of the two-point correlation function}\label{sec:xishape}

\begin{figure}
\begin{center}
{\epsfxsize=8.5truecm
\epsfbox[9 16 232 284]{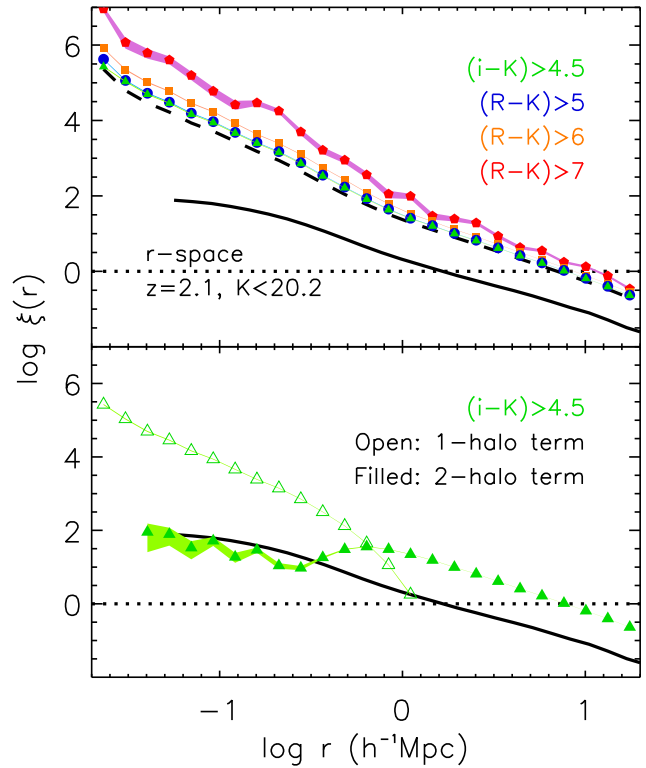}}
\caption
{The predicted real-space two-point correlation function, $\xi$, at $z=2.1$,
  as a function of comoving separation. The upper panel shows
  the predicted $\xi$ for all K-selected galaxies with $K< 20.2$ (dashed line), and
  for those which are also EROs, with  $(i-K)>4.5$ (triangles), $(R-K)>5$ (circles), $6$ (squares),
  $7$ (pentagons). The bottom panel shows the predicted $\xi$ for
  EROs with $(i-K)>4.5$, distinguishing between the contributions from the
  one-halo term (open triangles) and the two-halo term (filled
  triangles; see text for details). In
  both panels we also show the predicted $\xi$ of the dark matter (solid
  line). The dotted line shows $\xi=1$ and the
  shaded areas show the 1$\sigma$ Poisson errors derived using the
  number of pairs predicted by the model at a given separation. 
}
\label{fig:xi.cen_rk}
\end{center}
\end{figure}

\begin{figure}
\begin{center}
{\epsfxsize=8.5truecm
\epsfbox[13 7 238 209]{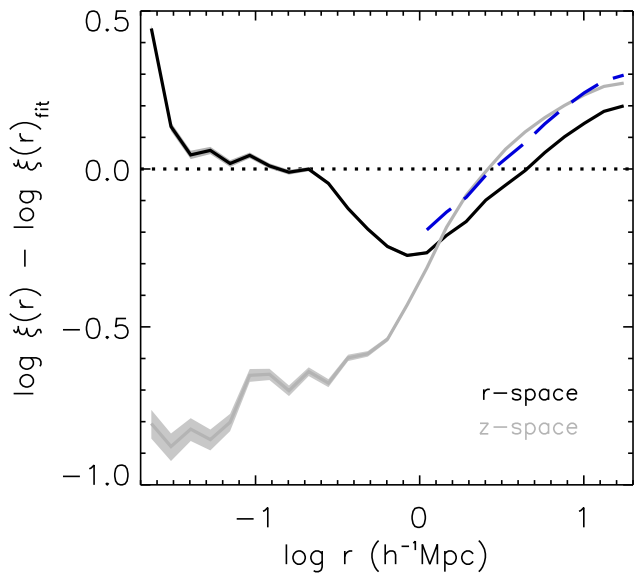}}
\caption
{The predicted ratio between the spatial correlation function of EROs, with
  $(R-K)>5$ and $K<20.2$ at $z=2.1$, and a power law fit,
  $\xi_{\rm fit}=(r/r_{0})^{\gamma}$, where $r_{0}$ is $7.0\, h^{-1}{\rm Mpc}$,
  and $\gamma$ is $-2.1$. The solid black (grey) line corresponds to
  the ratio using the real (redshift) space correlation functions. The dashed line is the
  redshift-space clustering for $r>1h^{-1}$Mpc obtained by multiplying the clustering in
  real-space by the Kaiser factor (Eq. \ref{eq:kaiser}, \S\ref{sec:kaiser}). The shaded areas show the 1$\sigma$ Poisson errors derived from the number of pairs at a given separation. 
}
\label{fig:norm_clustering}
\end{center}
\end{figure}

The predicted real-space two-point correlation function at $z=2.1$ for galaxies with $K<20.2$, including EROs, is plotted in
Fig. \ref{fig:xi.cen_rk}. This magnitude limit has been chosen
to match that used in deep observational surveys \citep{roche02,miyazaki03}. The
predicted shape
of the two-point correlation function for all K-selected galaxies and EROs is roughly consistent
with a power law for pair separations in the range $0.03\lesssim r\,
\lesssim 30 h^{-1}{\rm Mpc}$.

Fig. \ref{fig:norm_clustering} plots the ratio between the correlation
function of the EROs and a power
law, $\xi_{\rm fit}=(r/r_{0})^{\gamma}$, of EROs  with  $(R-K)>5$
and $K<20.2$ at $z=2.1$. This emphasizes any deviation from a power
law of the real-space correlation function. The parameters
of the power law  $\xi_{\rm fit}$ are those that give the best fit to
the real space correlation function
of the EROs plotted in Fig. \ref{fig:norm_clustering}: $r_{0}=7.0\,
h^{-1}{\rm Mpc}$ and $\gamma=-2.1$. Fig. \ref{fig:norm_clustering}
shows that there is a change in slope at
$r\sim 1 h^{-1}{\rm Mpc}$. We have quantified this change by fixing
$r_{0}$ and then fitting $\xi$ separately 
for $r<1 h^{-1}{\rm Mpc}$ and for larger separations. We have found a
change in slope $|\Delta\gamma|\leq0.5$ (less than $25$\%) between
large and small scales for the case of EROs
selected with different colour and K-magnitude cuts and at redshifts $z=1,\, 1.5$ and $2$. 

Gravitational instability theory predicts the existence of an
inflection point in the two-point correlation function of dark matter
when a smooth initial power spectrum is assumed, such as a power law. This occurs at the transition from the linear to the
non-linear regime, which is related to the one-halo to two-halo transition \citep[see e.g.][]{peebles80,gazta01,scoccimarro01}. From both semi-analytical models
and smoothed particle hydrodynamics simulations, a similar change in slope
is expected to occur for the two-point correlation function of
galaxies, for separations close to the dark matter correlation length, $r_{0}$,
i.e. the separation for which $\xi_{DM}(r_0)=1$. The strength of the
inflection depends on the galaxy selection which determines the number
of galaxies per halo and the balance between the one and two halo
term. This transition is supported by observations at low redshift
\citep[e.g.][]{baugh96,gazta01,zehavi02}. We can see such behaviour in Figs. \ref{fig:xi.cen_rk} and \ref{fig:norm_clustering}. Around $1 h^{-1}{\rm Mpc}$, the
two-point correlation function measured for all K-selected galaxies
and EROs has a small change of slope that makes the correlation function steeper on small scales. 

On small scales, $r< 1 h^{-1}{\rm Mpc}$, $\xi$ mainly measures the number of
pairs of galaxies within the same halo, i.e. the one-halo
term \citep[e.g.][]{benson00}. The
one-halo term is dominated by the distribution of satellites within single
haloes. Thus, a sample with a higher
proportion of satellites will have a boost in the
clustering on small scales. The distribution of satellite galaxies
depends strongly on the selection criteria (magnitude limit, colour
cut, redshift range, etc.). The lower panel of
Fig. \ref{fig:xi.cen_rk} shows separately the contribution of the
one-halo and two-halo terms to the predicted global clustering of EROs
with $(i-K)>4.5$. We note that the two predicted contributions for EROs with $(R-K)>5$ are
almost indistinguishable from those for EROs with $(i-K)>4.5$. 

On larger scales, $r> 1\, h^{-1}{\rm Mpc}$, we are generally probing
the positions of galaxies in different halos, the two-halo
term. We can see in the lower panel of Fig. \ref{fig:xi.cen_rk} that
the two-halo term of EROs has the same shape as the correlation
function of the dark matter. On these scales, the two point
correlation function of galaxies traces that of their host dark matter
haloes scaled by their bias and the dark matter clustering is close to the
prediction from linear theory on these scales.  We measure the
  clustering of EROs with $(R-K)>5$ and $K<20.2$ at $z=2.1$, $\xi_{gg}(r)$, to be boosted with respect to that of
dark matter, $\xi_{DM}$, with a bias $b\simeq
\sqrt{\xi_{gg}(r)/\xi_{DM}}\sim 3$ (at $r=8h^{-1}$Mpc). This is
consistent with these galaxies being hosted by haloes with
$M_{halo}\sim 4\times 10^{13}h^{-1}M_{\odot}$, in agreement with the predictions shown in
Fig. \ref{fig:hod}. 

In the lower panel of Fig. \ref{fig:xi.cen_rk} it is noticeable that the two-halo term
for EROs does not go to zero at scales where the
one-halo term dominates. For EROs there is
a large satellite component. It is then possible for two haloes that
are almost touching, to have satellite pairs at a smaller
separation than that corresponding to the centres of the two haloes. 

\subsubsection{Variation of clustering with colour}

Fig. \ref{fig:xi.cen_rk} shows that at $z=2.1$ galaxies in the ERO
samples defined by the bluest cuts, with $(R-K)>5$ or
$(i-K)>4.5$, cluster in a very similar way. In line with the results
described in \S\ref{sec:hod}, the clustering of these
bluest EROs is also seen to be very close to that of galaxies
selected solely on the basis of their K-band apparent magnitude without a cut
in colour. EROs with $(R-K)>5$ or
$(i-K)>4.5$ account for most of the bright end of
the luminosity function of K-selected galaxies at $z=2.1$, which
explains their similar clustering. 

Fig. \ref{fig:xi.cen_rk} also shows that the reddest EROs
display the strongest change in power law slope with pair separation. As discussed above, a larger proportion of satellite
galaxies is found among the redder EROs, explaining the boost seen at
small scales. 

\subsubsection{Variation of clustering with redshift}

\begin{figure}
\begin{center}
{\epsfxsize=8.3truecm
\epsfbox[23 8 238 208]{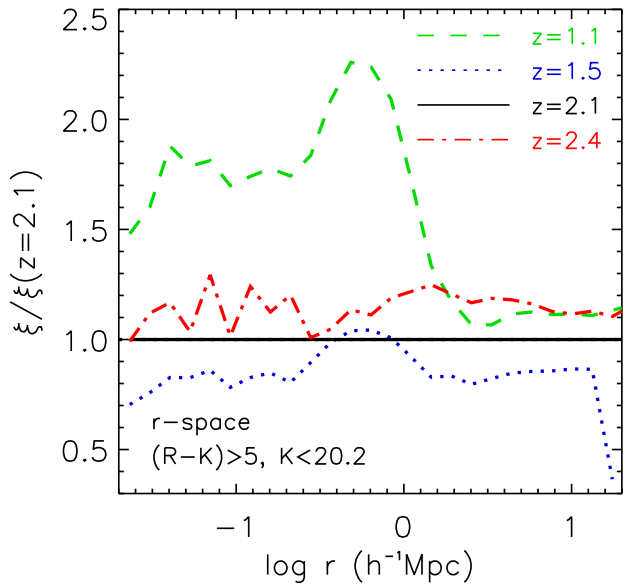}}
\caption
{Variation with redshift of the predicted real-space two-point correlation
  function for EROs with $(R-K)>5$  and $K<20.2$. Correlation
  functions at $z=1.1,\, 1.5,\, 2.1,\, 2.4$ (dashed, dotted, solid and dash-dotted line, respectively) are
  shown divided by the correlation function predicted at $z=2.1$. 
}
\label{fig:xi.ev_z}
\end{center}
\end{figure}

\begin{table}
\caption{
{\small The predicted real-space comoving correlation length, $r_0$, for EROs with $(R-K)>5$
and different magnitude cuts, at $z=1.1,\, 1.5,\, 2.1$. Not enough
EROs with $K<18.4$ are found at $z=2.1$ to have an estimation of their
clustering.}}
\begin{center}
\begin{tabular}{cccc}
\hline
 &   &  $r_{0}(h^{-1}{\rm Mpc})$ &  \\
\cline{2-4}
$z$   & $K<20.2$    & $K<19.2$   & $K<18.4$  \\
\hline
 1.1  & 8.4  & 8.1 &  7.5 \\
 1.5  & 7.2  & 7.3 &  7.2 \\
 2.1  & 7.9  & 7.3 &   -  \\
 \hline
\end{tabular}\label{tab:r0}
\end{center}
\end{table}

Table \ref{tab:r0} presents the real-space comoving correlation length, $r_{0}$, of EROs
with $(R-K)>5$ for different redshifts and magnitude ranges. The correlation length is
obtained here as the galaxy separation that gives $\xi(r_0)=1$.

In Table \ref{tab:r0} we can see that $r_{0}$ does not change
monotonically with redshift. This lack of a clear evolutionary
trend has also been found observationally by \citet{brown05}. A
similar prediction was made from the theoretical predictions in \citet{baugh99}
for a sample with a fixed apparent magnitude limit, so this trend is
not necessarily peculiar to the colour selection.

The variation of the clustering of EROs with redshift is
small. Therefore, we have emphasized this variation by presenting in Fig. \ref{fig:xi.ev_z} the clustering of EROs
with $(R-K)>5$ and $K<20.2$ at different redshifts, $z=1.1, \, 1.5,\, 2.1$ and
$2.4$, divided by that at $z=2.1$. It is clear from
Fig. \ref{fig:xi.ev_z} that the two-point
correlation function of EROs in real-space presents a slightly different shape
at $z=1.1$ than at higher redshifts, showing a boost on small scales
$r\lesssim 0.5 h^{-1}{\rm Mpc}$. This is related to the higher
fraction of satellite galaxies selected at $z=1.1$ as EROs, in
comparison with the higher redshifts plotted, as reported in \S\ref{sec:hod}.

The evolution of the correlation function with redshift is important
for interpreting the angular clustering of galaxies. This evolution is often parametrised as a
global, scale independent factor in redshift.
Fig. \ref{fig:xi.ev_z} shows that the variation of the
clustering with redshift is different on large and small scales. This
makes it difficult to assign a unique redshift power law, $1/(1+z)^{\alpha}$, to
describe this evolution, so such ``deprojections'' of angular
clustering are at best approximate.

\subsubsection{Variation of clustering with apparent magnitude}

\begin{figure}
\begin{center}
{\epsfxsize=8.5truecm
\epsfbox[9 14 234 288]{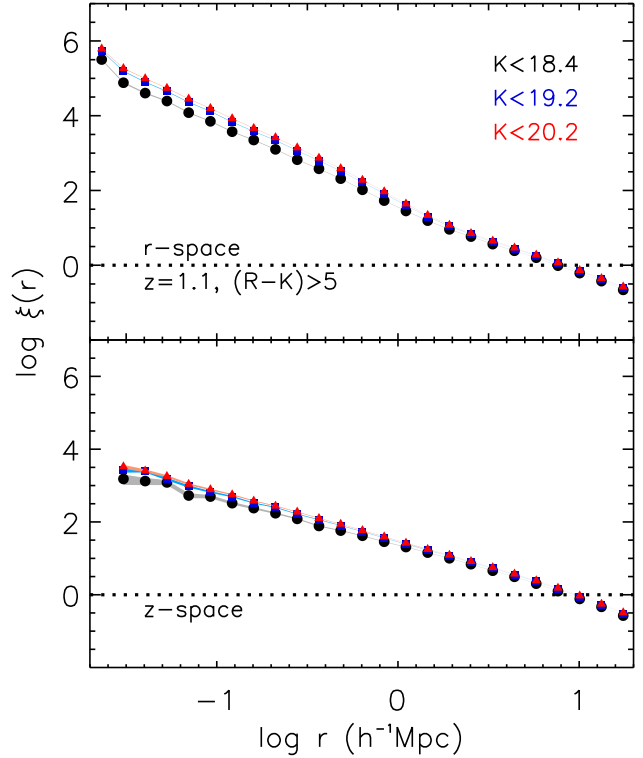}}
\caption
{The predicted spatial two-point correlation function of EROs, with $(R-K)>5$,
  brighter than $K<18.4$ (circles), $19.2$ (squares), $20.2$
  (triangles) at $z=1.1$. The real-space prediction is shown in the
  upper panel and the redshift-space in the bottom. In both
  cases, the shaded areas show the 1$\sigma$ Poisson errors,
  derived using the number of pairs predicted at a
  given comoving separation.
}
\label{fig:xi.ev_mag}
\end{center}
\end{figure}

Fig. \ref{fig:xi.ev_mag} shows the predicted two-point correlation
function at $z=1.1$ in both real and redshift-space for EROs with $(R-K)>5$ and
$K<18.4,\, 19.2$ and $20.2$. At this redshift, the model predicts the existence of EROs in sufficient
numbers for each of these magnitude cuts to allow as to compare the clustering
characteristics. We find that the clustering of EROs depends weakly
on luminosity. Fainter EROs are predicted to be slightly more clustered than brighter
ones, due to a higher fraction of satellite galaxies in massive
haloes. Fig. \ref{fig:xi.ev_mag} shows that the strongest
variation occurs on small scales. Table \ref{tab:r0} also shows a weak trend for fainter galaxies
to have a higher comoving correlation length, $r_{0}$.

Table \ref{tab:r0} lists the predicted $r_{0}$ for EROs selected in magnitude
ranges similar to various observational studies that inferred
the correlation length by fitting models to the measured angular correlation functions.
\citet{daddi01} analysed $400$ EROs with $K_{S}<19.2$, $(R-K_{S})>5$
and a median redshift $z\sim 1.2$ in a $701\, {\rm arcmin}^2$ field and estimated that their correlation length is not less than $r_0=8h^{-1}$Mpc, with a most
probable estimate of $r_0\sim12\pm 3h^{-1}{\rm Mpc}$. \citet{roche02} inferred that EROs with $K<20.2$, $(R-K)>5$
and $1\leq z_{\rm photo}\leq 3$
($158$ EROs in a $81.5\, {\rm arcmin}^2$ field), have a correlation length
between $10h^{-1}$Mpc and $13h^{-1}$Mpc. \citet{brown05} inferred the correlation length of EROs with $K<18.4$,
$(R-K)>5$ and a median photometric redshift of $z \sim 1.2$ ($671$
EROs in a $3529\, {\rm arcmin}^2$ field), to be
$9.7\pm1.0\,h^{-1}{\rm Mpc}$. 

Table \ref{tab:r0} shows that the predicted correlation length for
EROs is consistently lower than that estimated from the
observations. However, the predicted correlation length for EROs with
$K<19.2$ at $z=1.1$ is above the lower limit estimated by
\citet{daddi01} and within the $2\sigma$ range of their best
estimate of $r_{0}$. The comparison
with correlation lengths deduced from observations is not ideal since
a number of assumptions have to be made to infer this quantity from the actual
observables. In particular, all the observational studies mentioned above
assumed a power law for the spatial two-point correlation function, $\xi$, with
a fixed index $\gamma \approx -1.8$. The predicted $\xi$ is best fitted by
a power law with an index $\gamma \approx -2.1$ at $z=1.1$ and at
$z=2.1$, as shown at the beginning of \S\ref{sec:xishape}. Thus, if
this predicted value was adopted for the observational estimation, lower
correlation lengths would be obtained, resulting in a better agreement
between the model and observations.
A direct comparison with the observed two-point angular
correlation function will be carried out in \S\ref{sec:w}.

\subsubsection{Clustering in real and redshift-space}\label{sec:kaiser}

To mimic a spectroscopic survey in which radial positions of galaxies
are inferred from their redshifts, galaxy positions are perturbed
along one of the axes by their peculiar velocities, scaled by the appropriate value of the
Hubble parameter. The impact of these peculiar motions depends on the
scale. Figs. \ref{fig:norm_clustering} and \ref{fig:xi.ev_mag} show that while the slight
change of slope around $1h^{-1}$Mpc in the correlation function of EROs is clear in real
space, this is smeared out when including redshift-space distortions.
 
Fig. \ref{fig:norm_clustering} emphasizes the differences
between the correlation functions in real and redshift-space by plotting the
correlation functions divided by the power law that best fits the
real-space clustering of EROs. Both Figs. \ref{fig:norm_clustering} and \ref{fig:xi.ev_mag} show that
the difference between real and redshift-space clustering depends on
scale. 

On small scales, $r \lesssim 1 h^{-1}{\rm Mpc}$, the clustering in
redshift-space is significantly lower than that in real
space. On these scales we are generally considering galaxies within the same halo.  As can be seen in
Fig. \ref{fig:hod}, the \citeauthor{bower06} model predicts the
existence of more than one ERO in halos of high enough mass, except
for the most extreme EROs with $(R-K)>7$. The peculiar motions of
EROs within a halo cause an apparent stretching of the structure in
redshift-space, diluting the number of ERO pairs at small separations. Thus, the
correlation function signal measured in redshift-space is smaller
than that in real-space on these scales.

On larger scales, $1 \lesssim r\lesssim 30 h^{-1}{\rm Mpc}$, bulk motions of galaxies cause the galaxy distribution to appear squashed along the line of sight in
redshift-space, enhancing the clustering amplitude. Fig. \ref{fig:norm_clustering} shows that, on this range of scales,
the boost measured for the redshift-space correlation function is in
reasonable agreement with the value expected according to the formalism developed by \citet{kaiser87}. This relates the
correlation function in redshift-space, $\xi_{gg}(s)$, to that in
real-space, $\xi_{gg}(r)$,
through a factor, $f$, $\xi_{gg}(s)=f\xi_{gg}(r)$. This factor is a
function of the bias, $b$, and the matter content of the
Universe, $\Omega_m$:
\begin{equation}\label{eq:kaiser}
f= 1+\frac{2}{3}\frac{\Omega_m^{\gamma '}}{b}+\frac{1}{5}\Big(\frac{\Omega_m^{\gamma '}}{b}\Big)^2 
\end{equation}
In fact, the redshift-space clustering is only expected to be
accurately described by the ``Kaiser factor'' on very large scales,
since linear perturbation theory is assumed
\citep{elise11}. Traditionally $\gamma '=0.6$ but $\gamma '=0.55$ is a
better approximation \citep{linder05} and is the value we have used here.

\subsection{Clustering of quiescent and star forming EROs}\label{sec:qui}

\begin{figure}
\begin{center}
{\epsfxsize=8.5truecm
\epsfbox[32 8 238 196]{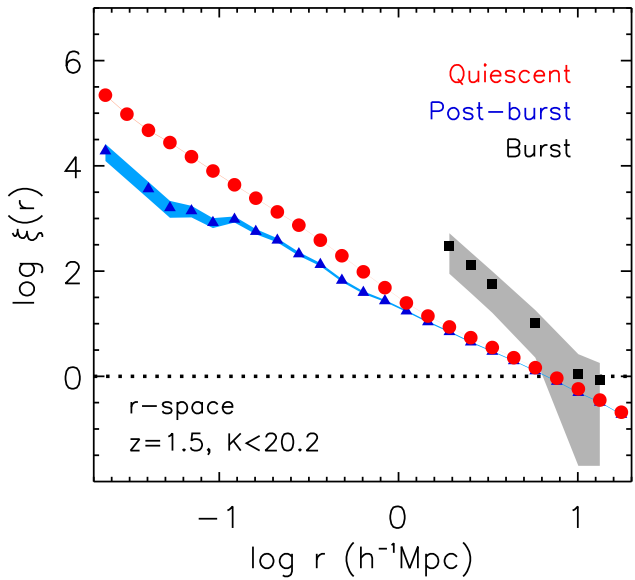}}
\caption
{The predicted real-space two-point
  correlation function for EROs with  $(R-K)>5$ and $K< 20.2$ at
  $z=1.5$, separated into {\it burst} (squares), {\it
  post-burst} (triangles) and {\it
    quiescent} (circles) galaxies (see text in \S\ref{sec:qui}  for details).
}
\label{fig:xi.ev_sfr}
\end{center}
\end{figure}

%
%

Observationally, EROs exhibit a mixture of spectral properties which
are related to their star formation histories and dust content. In
this section we analyse the differences in the predicted clustering between EROs that are actively forming stars and those
that are passively evolving.

Fig. \ref{fig:xi.ev_sfr} shows the clustering of EROs with
$(R-K)>5$ and $K<20.2$ at $z=1.5$, divided into {\it burst} (galaxies
with an ongoing burst of star formation), {\it
  post-burst} (last burst happened within the past 1 Gyr) and {\it
  quiescent} galaxies (all the others). As described in
\citetalias{eros1}, this classification is made by considering the lookback time to the start of the last burst of star formation
experienced by the galaxy. In the \citeauthor{bower06}
model bursts are triggered by galaxy mergers and by disks becoming
dynamically unstable.

The higher clustering amplitude predicted in real-space for {\it
  burst} EROs, $r_{0}=11.7 h^{-1}{\rm Mpc}$, is noticeable compared
with that of {\it quiescent} ones, for which $r_{0}=7.3 h^{-1}{\rm Mpc}$,
and for {\it post-burst} ones, with $r_{0}=6.7
h^{-1}{\rm Mpc}$. However, Poisson errors dominate the {\it burst}
ERO clustering estimate, due to the much lower space density of these EROs
compared with the other two samples. We can see in
Fig. \ref{fig:xi.ev_sfr} how the correlation functions for 
{\it quiescent} and {\it post-burst} EROs are similar for
separations larger than $r\sim 1 h^{-1}{\rm Mpc}$, with slopes differing by
$|\Delta\gamma|<0.2$ on small scales. The same tendencies are found in redshift-space. 

In \citetalias{eros1}, we explored the predicted mix of quiescent and
active EROs using different classifications. In particular, we used the colour-colour diagram
proposed by \citet{pozzetti00}, the specific star formation rate of
EROs, adopting $SFR/M_{*}=10^{-11}{\rm yr}^{-1}$ as the boundary\footnote{
At $z=1.5$, the inverse of the age of the
Universe is $10^{-9.6}{\rm yr}^{-1}$, however there are very few EROs with
$SFR/M_{*}>10^{-10}{\rm yr}^{-1}$. The specific star
formation boundary value was chosen in  \citetalias{eros1} taking into account both the
predicted distribution and number counts of EROs.
}, and the classification described above, but considering as active
both {\it post-burst} and {\it burst} EROs. We found that the
number counts of quiescent EROs agreed among the different
classifications. $75$\% of the quiescent EROs in Fig. \ref{fig:xi.ev_sfr} have
$SFR/M_{*}<10^{-11}{\rm yr}^{-1}$. We find that the
EROs with $SFR/M_{*}<10^{-11}{\rm yr}^{-1}$ have a correlation length $r_{0}=7.7
h^{-1}{\rm Mpc}$ and that, globally, their two-point correlation function is very close to
that for {\it quiescent} EROs. The correlation lenght of EROs with
$SFR/M_{*}>10^{-11}{\rm yr}^{-1}$ is predicted to be $r_{0}=5.7h^{-1}{\rm
  Mpc}$, slightly lower than that for  {\it post-burst} EROs.

The active EROs, {\it post-burst} plus {\it burst} EROs, are dominated
in our model by the {\it post-burst} galaxies. The predicted clustering of active EROs practically overlaps with that for {\it
  post-burst} EROs only, having the same predicted correlation length
$r_0= 6.7 h^{-1}{\rm Mpc}$.

A few observational studies have tried to obtain the real-space
correlation length differentiating between quiescent EROs and
dusty star forming or active EROs. Among these are those by \citet{daddi02} and \citet{miyazaki03}
studies, which inferred the correlation length of EROs from the
measured angular clustering.

\citet{miyazaki03} observed EROs with $(R-K_S)>5$ and $K_S<20.2$, dividing the sample into quiescent and star forming galaxies
by fitting observed broad band colours to synthetic spectral energy distributions. Without taking into account the photometric
redshift errors, \citeauthor{miyazaki03} found the correlation functions of
quiescent and active EROs to be $r_{0}=11\,\pm\, 1\,h^{-1}$Mpc and
$r_{0}=12\,\pm\, 2\,h^{-1}$Mpc, respectively. In agreement with the observations of  \citeauthor{miyazaki03}, we find that quiescent and active EROs have
similar correlation lengths. Taking into account the photometric redshift errors,
\citeauthor{miyazaki03} revised their estimate of the correlation length for quiescent EROs to be $r_{0}=8
h^{-1}$Mpc. This value is comparable to but higher than the predicted
correlation length of quiescent EROs with $(R-K)>5$ and
$K<20.2$ at $z=1.5$.  

\citet{daddi02} observed EROs
with $(R-K_S)>5$ and $K_S<19.2$, splitting the sample using spectral
features. \citeauthor{daddi02} estimated the correlation length to be $r_{0}<2.5\, h^{-1}{\rm Mpc}$ for star forming EROs and $5.5 \lesssim
r_{0} \lesssim 16\, h^{-1}{\rm Mpc}$ for quiescent EROs. Our
predicted $r_{0}$ values for quiescent EROs agree within errors with
those from \citeauthor{daddi02}. The correlation function for
quiescent EROs inferred by \citet{miyazaki03} agrees with the estimate by \citet{daddi02}, though they used
different magnitude cuts. However, these two observational studies
estimated very different correlation lengths for star forming EROs. Sample variance could be
responsible, at least partly, for the difference between the observational studies, since
\citeauthor{daddi02} covered a field $7$
times larger than that used in the \citeauthor{miyazaki03} study. Another
issue that will increase the discrepancy is the different depths of
the two observations. \cite{cimatti02a} observed that star forming EROs
appear at slightly higher redshifts than old EROs. This implies that
surveys sampling only bright galaxies will contain a lower percentage of star
forming EROs. The need for more fields to be
explored observationally is clear, in order to have a better
knowledge of the origin of such different $r_{0}$ values for active EROs.


\section{The two-point angular correlation function of EROs}\label{sec:w}

In order to make a direct comparison with current measurements
of clustering, we study here the predicted angular correlation function
$\omega(\theta)$ for EROs, obtained from our previous estimate of
the two-point spatial correlation function in real space and the
redshift distribution of model EROs.

The Limber equation \citep{limbereq}
relates the spatial correlation function $\xi$ to the angular
correlation function $\omega$, under the assumptions:
\begin{enumerate}
\item We can apply a sample selection and measure the correlation
  function for that full sample, without any concern for dependencies
  of clustering on intrinsic galaxy properties which vary within the
  sample \citep[e.g.][]{peebles80}.
\item The selection function does not vary over the pair separations on which we can
measure a signal for $\xi$ \citep[see e.g.][]{simon07}. For most
surveys this is a reasonable assumption, since the signal for $\xi$
will be measurable only for those pairs of galaxies such that their pair
separations $|r_1-r_2|$ are much smaller than their comoving distances, $r_1$, $r_2$.
\end{enumerate}
The above approximations are both reasonable for EROs. Therefore, for a flat
Universe
\footnote{For open cosmologies see the general expression given by \citet{baugh93}.} 
we can calculate the angular correlation function, $\omega$,
for pairs of EROs separated by an angle $\theta$ by:
 \begin{equation}\label{eq:limber}
\omega ( \theta)= \frac{
2\int_0^{\infty}\Big[ \frac{{\rm d}N}{{\rm d}z} \Big]^2 \frac{{\rm d}z}{{\rm d}r} \, 
\Big(  \int_0^{2r}{\rm d} u \, \xi(r_{12},z) \Big) \, {\rm d} z
}{
\Big[ \int_{0}^{\infty} \frac{{\rm d}N}{{\rm d}z}\, {\rm d} z  \Big] ^2
} \, ,
\end{equation}
where d$N/{\rm d}z$ is the redshift distribution of the surveyed galaxies. For a pair of galaxies at comoving distances $r_1$ and  $r_2$:
$u=r_1-r_2$, $r=(r_1+r_2)/2$ and the comoving separation between them is approximated
by $r_{12}=\sqrt{u^2+r^2\varpi^2}$, with
$\varpi^2=2({\rm cos}\theta-1)$. The variation of redshift with
comoving distance d$z/{\rm d}r=(H_0/c)\sqrt{\Omega_{0}(1+z)^3+(1-\Omega_{0})}$, where $H_0$ is the Hubble constant now and $c$ is the
speed of light. When integrating the two-point correlation function
at very large scales, beyond those modelled accurately within the
simulation volume, we assume that on these scales $\xi$ is given by a scaled dark matter two-point correlation function
calculated from the linear initial power spectrum of density
fluctuations, where the scaling is the linear bias factor, $b^2$
\citep[see][]{orsi08}.

\subsection{EROs selected on (R-K) colour}\label{sec:xirk}
\begin{figure}
\begin{center}
{\epsfxsize=8.5truecm
\epsfbox[17 37 257 410]{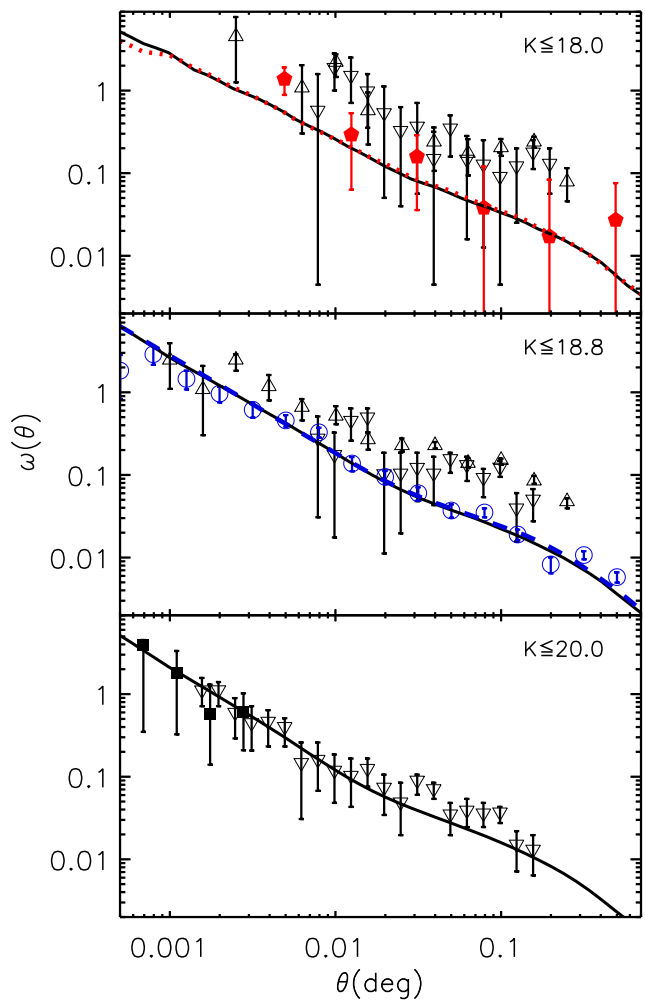}}
\caption
{The two-point angular correlation function of EROs with
  $(R-K)>5$ and brighter than $K=18,\, 18.8, \, 20$, from top to bottom. The black solid lines show the
  predicted angular correlation. The top panel also contains a dotted line
  with the clustering predicted for EROs with $K\le 17.9$. The middle
  panel includes a dashed line showing the predicted angular clustering for EROs with
  $(r-K)>5$ and $K\leq 18.8$. In
  each panel there are observational results for the corresponding
  EROs: \citet{daddi00} (triangles), \citet{roche02} (filled squares), \citet{brown05} (filled
  pentagons), \citet{kong06} (upside down triangles), \citet{kim11}
  (circles). Error bars are Poisson error bars, except for the data
  points from \citet{brown05}, which include the effects of sample
  variance and shot noise. 
}
\label{fig:w}
\end{center}
\end{figure}
 
Fig. \ref{fig:w} shows the predicted $w(\theta)$ for EROs with
$(R-K)>5$ and different apparent magnitude cuts, compared with
observations. 
The triangles in Fig. \ref{fig:w} show the clustering of EROs
studied by \citet{daddi00} in a field, Daddi-F, which covers $701\,
{\rm arcmin}^2$ and is $85$\%  complete to $K_S<18.8$. The upside
down triangles in Fig. \ref{fig:w} show the clustering of EROs studied
by \citet{kong06} in both a subsection of the Daddi-F field of $600\, {\rm
  arcmin}^2$ and in the Deep3a-F field with an area of $320\, {\rm arcmin}^2$. 
The squares in Fig. \ref{fig:w} show the clustering of
EROs measured by \citet{roche02} in a $81.5\, {\rm arcmin}^2$ field.  
The pentagons in Fig. \ref{fig:w} show the clustering of EROs with
$K<17.9$ measured by \citet{brown05} in a $0.98\, {\rm deg}^2$ field;
their error estimate uses a Gaussian approximation to the covariance matrix and, thus,
includes the effects of sample variance and shot noise. 
\citet{kim11} measured the clustering
of EROs with $(r-K)>5$ and $K\leq 18.8$ observed in the SA22 field with an area of
$2.45\, {\rm deg}^2$, one of the largest fields used to date to
measure ERO clustering. The \citet{kim11} data points are shown as open circles in Fig. \ref{fig:w}

Both  \citet{daddi00} and
\citet{kong06} observed roughly the same field using different
R-filters. Their estimates of the clustering of EROs agree within the errors,
though \citet{kong06} measured a slightly higher clustering amplitude than \citeauthor{daddi00} 

The top panel in Fig. \ref{fig:w} shows the predicted angular
clustering of EROs for both $K\leq 18$, solid line, and $K\leq 17.9$,
dotted line, to match the observational samples. The predicted difference is minimal, as expected for such
a small change in magnitude. The top panel in Fig. \ref{fig:w} shows that the
model reproduces the clustering estimated by \citet{brown05} for EROs
with $K\le 17.9$ in a field $5$ times larger than that observed by \citet{daddi00} and
\citet{kong06} (Daddi-F field). 

The middle panel in Fig. \ref{fig:w} shows the predicted angular
clustering of EROs with $K\leq 18.8$ compared with the estimates
from \citet{daddi00}, \citet{kong06} and \citet{kim11}. As we found for
brighter EROs, our predicted angular clustering is lower than that
estimated by  both \citeauthor{daddi00} and
\citeauthor{kong06} (Deep3a-F field). However, it is clear from Fig. \ref{fig:w} that the model
predictions match the clustering for EROs with $K\le 18.8$ estimated by
the UKIDSS team \citep{kim11} over an area $12$ times larger
than the Daddi-F field. The middle panel in Fig. \ref{fig:w} also shows the predicted
angular clustering for EROs selected with $(r-K)>5$, the same filters 
used by \citet{kim11}. Both $r$ and $R$ filters are very close, and
the difference in the predicted angular clustering is negligible.

For EROs with $(R-K)>5$ and $K\le 20$, the bottom panel of Fig. \ref{fig:w} shows that the predicted clustering reproduces the
observational estimates from both \citet{roche02} and
\citet{kong06} (Deep3a-F field). This is also true for EROs with $K\le 19.5$.

\citet{kong06} found an increase in the amplitude of the angular correlation
function with the brightness of EROs. The observed variation of clustering with magnitude
is stronger than that predicted, even after taking into account some
small discrepancies due to the use of slightly different filters. This
is not peculiar to EROs, since a
similar conclusion was reached by \citet{kim10}, who studied bright
galaxies at $z=0$, finding that the clustering predicted by semi-analytical
models does not vary with luminosity as strongly as observed. \citet{kim10} concluded that a stronger dependence of clustering
strength on luminosity could be obtained by including additional
physical processes which influence the mass of satellite
galaxies. This could have an implication for EROs as an appreciable
fraction of our EROs are satellites.

Nevertheless, once the sampling variance is taken into account, the
model predictions do match the observed angular clustering reasonably
well for $0.0006\le \, \theta ({\rm deg})\, \le 0.6$.

\subsection{EROs selected on (i-K) colour}\label{sec:xiik}

Accurate measurements of the clustering of EROs can only be achieved by 
deep imaging of large areas. Currently, the UKIDSS survey
\citep{kim11} is one of the few surveys that has been able to measure the
clustering of EROs beyond separations of $0.1$ deg (which corresponds
to $\sim 2\,h^{-1}{\rm Mpc}$ at $z=1.5$), which probes the two-halo component of the correlation function. In
this section we show the model predictions for EROs selected in a
similar way to that adopted by \citet{kim11}, i.e. using the $(i-K)>4.5$
colour criterion. This criterion appears to select galaxies with $z\ge
1$, with fewer lower redshift contaminants than the $(R-K)$ colour criterion \citep{kong09}.

\subsubsection{Number counts and redshift distribution}
\begin{figure}
\begin{center}
{\epsfxsize=8.5truecm
\epsfbox[49 7 546 474]{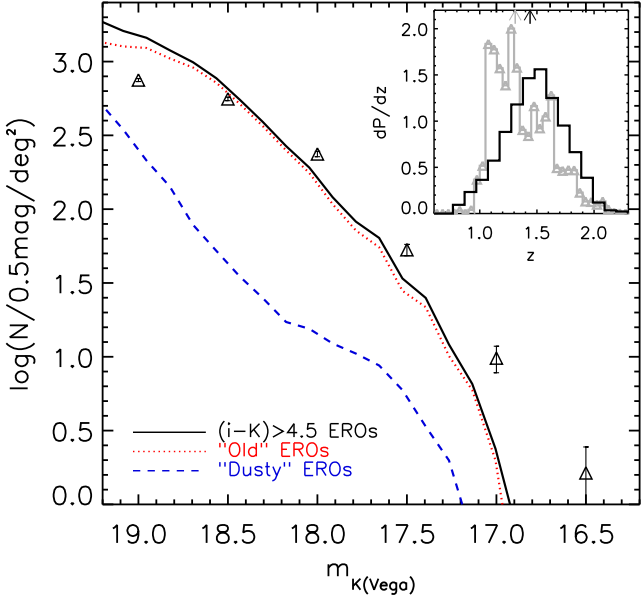}}
\caption
{Differential number counts of EROs with $(i-K)>4.5$. The predicted
  number counts for all EROs are shown by a solid line. The
  prediction for ``old" (``dusty") EROs are presented with dotted (dashed)
  lines (see \S\ref{sec:olddusty} for details on this division). The
  triangles show the observations from \citet{kim11}, which are
  complete up to $K\le 18.8$. The
  inset shows the redshift distribution for EROs with $(i-K)>4.5$ and
  $K \le 18.8$: predictions are in black and observations in grey. The
  areas of the histograms are normalised to unity. The median redshifts are
  shown by the vertical arrows. 
}
\label{fig:iknc}
\end{center}
\end{figure}

The success of a model in reproducing the observed angular clustering
of a certain type of galaxy is more robust if the model also reproduces their number density. Here we explore the predicted number
counts and redshift distributions for EROs selected by their $(i-K)$
colours. We already showed the predicted
abundance and redshift distributions for $(R-K)$ selected EROs in \citetalias{eros1}.

Fig. \ref{fig:iknc} shows the predicted differential number counts for
EROs with $(i-K)>4.5$, compared with the observations from
\citet{kim11}. These observations are complete for galaxies with $K \le
18.8$ \citep[see][for further details]{kim11}. The
model underpredicts the number counts at the bright end, but gives a
good match from $K=18$ to the completeness limit. 

The inset in Fig. \ref{fig:iknc} shows the predicted redshift
distribution of EROs with $(i-K)>4.5$ and $K \le 18.8$. For
comparison we have included, in grey, the redshift distribution
estimated by \citet{kim11}. This estimate  is based on
photometric redshifts measured by the NEWFIRM Medium Band Survey
\citep{brammer09,vandokkum10}. The area sampled by the NEWFIRM survey is
rather small, $0.25\, {\rm deg}^2$, and so is subject to significant
sample variance, with the result that distinctive features appear in
the redshift distribution. The model predicts a
redshift distribution close to the observed one, but with
a slightly higher median redshift $z=1.44$, as opposed to a median of
$z=1.31$ for the observational data.

The model predicts a population of EROs defined with $(i-K)$ colours
that is close enough in number and redshift distribution to the
observations to make it worthwhile to continue
calculating their clustering properties. For this purpose, we will
follow the method described above for EROs selected based on their
$(R-K)$ colour.

\subsubsection{Angular clustering}
\begin{figure}
\begin{center}
{\epsfxsize=8.5truecm
\epsfbox[19 19 256 283]{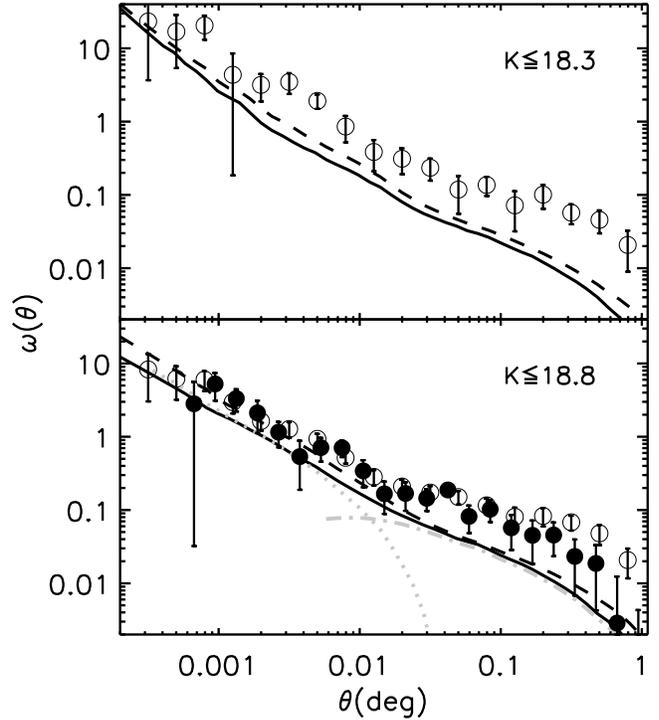}}
\caption
{The predicted two-point angular correlation function of EROs with
  $(i-K)>4.5$ and $K\le 18.3$ in the top panel, $K\le 18.8$ in the 
  bottom panel. The solid black lines show the prediction from the
  model. The dashed black lines show the model predictions using the
  observationally estimated redshift distribution from
  \citet{kim11}. The grey lines in the bottom panel present
  the predicted contribution of the one halo (dotted line) and two halo
  terms (dash-dotted line). Observational results from \citet{kim11} for the
  UKIDSS DXS SA22 field are shown as
  open circles. The filled circles present the
  observational estimation for the UKIDSS Elais-N1
  field \citep{kim12}. Observational data points are shown with Jackknife error bars. 
}
\label{fig:wik}
\end{center}
\end{figure}

Fig. \ref{fig:wik} shows the predicted two-point angular correlation function of EROs with
  $(i-K)>4.5$ compared with the observed angular correlation function measured within two UKIDSS
fields: SA22 \citep{kim11}, open circles, with an area of $2.45\,
  {\rm deg}^2$ and Elais-N1,
filled circles  \citep{kim12} with an area of $3.88\,
  {\rm deg}^2$ (both quoted areas are after masking). The SA22 field
  was observed with both the CTIO and UKIRT telescopes, while the Elais-N1
  field was observed with the Subaru and UKIRT telescopes. The observed correlation
  function in both fields is corrected for the integral constraint. For
  the Elais-N1 field the integral constraint is estimated to be
  $0.0073$, which is comparable to
  the clustering signal at scales of $\approx 0.7$ deg. For the clustering in both fields Jackknife errors have been
estimated following \citet{sawangwit09}, in an attempt to take into account the sampling variance \citep{norberg09}. 

We predict the angular clustering using Limber's
equation (Eq. \ref{eq:limber}), assuming either the predicted (solid lines) or the observed (dashed lines) redshift
distribution for the galaxies under study. We can see from
Fig. \ref{fig:wik} that the difference between using the predicted and observed
redshift distributions is negligible. The evolution with redshift of
the spatial two-point correlation function for EROs is modest, which
explains the small variation seen for the angular clustering when
using a different redshift distribution.

From the \citet{kim11} observations, the correlation function of EROs selected with
$(i-K) >4.5$ appears to be better described by two power laws, rather
than one, with the change of slope happening at $\theta\sim 0.02\,
{\rm deg}$. The model also predicts
such a change in slope, though this is milder than observed. The bottom panel in Fig. \ref{fig:wik} shows
how this change in slope is due to the change from
counting pairs of galaxies within the same halo, the one halo term, to
counting those from different haloes, the two halo term, as was also
seen for the spatial correlation function in \S\ref{sec:xi}. At scales
larger than $\theta\ge 0.2$ deg we can see the change in
the curvature of $\omega(\theta)$ intrinsic to the two-halo term, that follows the shape of
the dark matter clustering.

Fig. \ref{fig:wik} shows that the model underpredicts the angular
clustering with respect to observations. The difference becomes larger
at larger pair separations beyond $\theta=0.02\, {\rm deg}$. However, as can be seen in
Fig. \ref{fig:wik}, the uncertainty due to sampling variance also increases
with pair separation, where the clustering
measured in the two fields shows the largest differences.

As indicated before, the differences between model predictions
and observations are not due to the differences in the redshift
distributions. Sampling variance could be behind the
discrepancy between observations and model predictions, at least at
small scales, $\theta<0.02\, {\rm deg}$. On larger scales the difference is
clear even when allowing for the sampling variance errors in
observations of different fields. On large scales, $\theta>0.02\, {\rm
  deg}$, the angular clustering is underpredicted by a factor of 3 for EROs with $K\leq
18.8$. 

The match between predictions and observations is worse
for EROs selected with $(i-K)$ colour than it is for $(R-K)$
colours. The photometric errors in the CTIO i-filter are larger than
those for the r-filter, which implies that the selection made
using $(i-K)$ is slightly noisier than with $(r-K)$, at least for the
SA22 field. In fact, the prediction for EROs with $(i-K)>4.5$ perfectly matches the angular
clustering estimated from observations of EROs selected with
$(i-K)>4$, potentially pointing towards a problem specific to the colour
cut. The predicted angular clustering using the Subaru i-band is
almost identical to that shown in Fig. \ref{fig:wik}. We
have further explored the origin of this discrepancy by predicting the
angular clustering of EROs with $(i-K)>5$. We find a similar trend
to that reported for the spatial clustering: the clustering of
redder EROs is boosted, with the larger variation seen at small scales. In line
with this prediction, both \citet{daddi00} and \citet{kim11}
observed that redder EROs have angular correlation functions with
higher amplitudes.

Optical colours in the model are affected by the treatment of the
ram-pressure stripping in satellite galaxies \citep{font08}. When
selecting EROs by their $(i-K)$ or $(R-K)$ colours we could be enhancing their
numbers by selecting satellite galaxies that are redder than they
should be due to an
oversimplification of their gas physics in the \citeauthor{bower06} model. This would change the
amplitude of the angular clustering, particularly at small scales. We have investigated this point
by calculating the angular clustering of EROs using the
\citet{font08} model. This model is an extension to that of \citeauthor{bower06} which introduces a more realistic
treatment of the stripping of hot gas in satellite
galaxies. \citeauthor{font08} also double the stellar heavy element yield, improving the match of the
locus of the red sequence to observational data at $z\approx 0$.

Using the \citet{font08} model we find that the number counts of
EROs with $(i-K)>4.5$ are overpredicted for $K<18$ by around a factor
of two. The redshift
distribution of EROs predicted
by the \citeauthor{font08} model has a non negligible fraction of galaxies
at $z<1$, with a median redshift of $1.13$, below that observed. At a
given redshift, the spatial correlation function predicted by the
\citeauthor{font08} model is very close to that predicted from the \citeauthor{bower06} model, for EROs with either $(i-K)>4.5$  or $(R-K)>5$. However, the angular clustering
calculated from the \citeauthor{font08} model is predicted to be
boosted by at most a factor of $\sim 1.5$ at large scales,
when compared with that from the \citeauthor{bower06} model. Therefore, the \citeauthor{font08}
model does not reconcile the predictions with
the observational estimates. Since the \citeauthor{font08} model includes changes
in both the stellar yield and the treatment of hot gas stripping that
have opposite effects on galaxy colours, we have run the
\citeauthor{bower06} model changing only the stellar yield
 to double the original value. The results of doing this exercise are
 similar to using the full \citeauthor{font08} model. Both the redshift
distribution and number counts change upon changing the yield. These
changes are such that the angular clustering remains close to that
obtained with the original model. Thus, we find that a change in stellar
yield cannot account for the difference between the model predictions
and the observations, and leads to some predictions actually agreeing
worse with the observations.

\subsubsection{``Old'' and ``dusty'' EROs with $(i-K)>4.5$}\label{sec:olddusty}

The number counts for EROs separated into
``old" and ``dusty" galaxies are shown in Fig. \ref{fig:iknc}. This separation has been done according to the
location of the EROs in the colour-colour diagram
proposed by \citet{fang09}, where the line $(J-K)\, =\, 0.20(i-K)\,+\,
1.08$ is defined to set the boundary between these types. This selection is similar
to that of \citet{pozzetti00}, but is tuned to the filters used to
select EROs with $(i-K)$ colours. \citet{fang09} compared EROs selected
with both this colour-colour selection criterion and on the basis of
their spectra, finding that the two methods
agree reasonably well. Nevertheless, according to \citeauthor{fang09}, $\sim 33$\% of the ``old'' EROs
selected by the colour-colour method are actually young EROs when
looking to their spectra.

We find that about $90$\%
of the EROs with $K\le 19$ are ``old" galaxies according to the \citeauthor{fang09} colour criterion. We have tested that a small shift of about $6$\% in the
boundary proposed by \citeauthor{fang09}, $(J-K)\, =\, 0.20(i-K)\,+\,
1.02$, produces a similar change in the percentage of ``old" EROs of a
$5$\%. Thus, the colour-colour separation is reasonably robust for the
predicted EROs, since they do not preferentially populate the region close the colour-colour boundary.

\citet{kim11} observed that $60$\% of their EROs
are ``old" galaxies, a lower fraction than the model predicts. This
result is similar to the conclusion from \citetalias{eros1} on the nature of EROs selected using
$(R-K)$ colours, although, in the case of the $(i-K)$
selection the difference with observations is
larger. One possible reason for this discrepancy is that although {\sc galform} broadly reproduces the observed colour bimodality
observed for SDSS galaxies at $z=0$ \citep{juan09}, the model
predictions are different in detail from the observations. However, we
have found that when using either the \citet{font08} model or the
\citeauthor{bower06} model with twice the original yield value, about
$60$\% or $67$\%, respectively, of the predicted EROs with $K\approx18$ are ``old'', in good
agreement with observations. Therefore, the
difference in the split between ``dusty" and ``old" EROs appears to be
connected with the value of the yield in the \citeauthor{bower06}
model.

\begin{figure}
\begin{center}
{\epsfxsize=8.5truecm
\epsfbox[17 8 266 181]{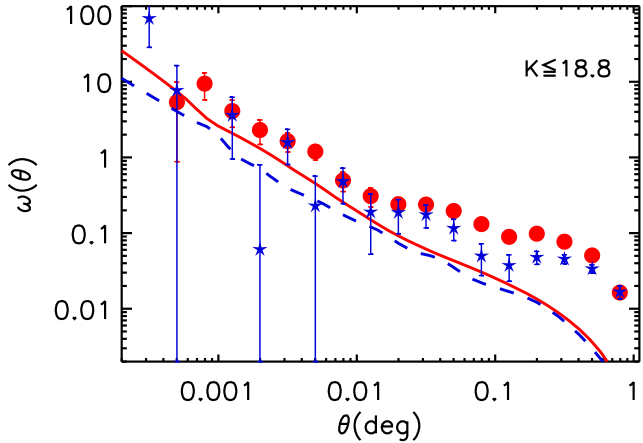}}
\caption
{The predicted two-point angular correlation function of ``old" (red
  solid line) and ``dusty" (blue dashed line) EROs with
  $(i-K)>4.5$ and $K=18.8$. The observational
  estimates of \citet{kim11} for ``old" (circles) and ``dusty" (stars) EROs are
  shown with Poisson error bars. See text for the details of the division
  between ``old" and ``dusty" EROs.
}
\label{fig:dgog}
\end{center}
\end{figure}


Fig. \ref{fig:iknc} shows that the percentage of ``old" EROs starts
to drop for fainter magnitude bins. The same tendency has been
observed by \citet{kim11}. \citet{kong09} also found the same trend for a sample of EROs selected with
$(I-K_S)>4$ and $K_S\le 19.2$.

The predicted median redshifts
for ``old" and ``dusty" EROs are $1.44$ and $1.25$,
respectively. \citet{kim11} estimated the median redshifts for the
observed ``old'' and ``dusty'' EROs to be  $z=1.36$ ($10$\%  and $90$\%
percentiles: $z=1.1$  and $z=1.7$, respectively) and
$z=1.25$ ($10$\%  and $90$\% percentiles: $z=1$  and $z=1.8$, respectively),
respectively, in reasonable agreement with the model predictions. Contrary to
these results, \citet{fang09} and \citet{kong09} found more ``dusty" EROs at higher
redshifts. However, we have found that this statement is strongly dependent
on the particular $(i-K)$ cut made, since when using $(i-K)>5$ we
find that ``dusty" EROs are predicted to appear at higher redshifts
than ``old" ones.

Fig. \ref{fig:dgog} shows the  predicted two-point angular correlation function of EROs with
 $(i-K)>4.5$ separated into ``old" and ``dusty" galaxies. Fig. \ref{fig:dgog}
 shows that the predicted clustering for both types of EROs differ
 only on small scales. Thus, model predictions and observations agree
in that both ``dusty" and ``old" EROs are clustered very similarly.

In the previous section we discussed the possible origin of the difference between the
predicted and the observed angular clustering for EROs with
$(i-K)>4.5$. The balance between ``dusty" and ``old" EROs when using
$(i-K)$ colours could alter the predicted angular clustering. However,
Fig. \ref{fig:dgog} shows that the clustering of ``dusty'' EROs is
both observed and predicted to be similar, though slightly less strong
than that for the ``old'' EROs. Therefore, a larger fraction of
``dusty'' EROs within the model would actually slightly increase the difference between
observations and model predictions, by further reducing the predicted angular clustering at large scales.

\section{Conclusions}\label{sec:conclusions}

In this paper we have extended the tests of the {\sc galform} galaxy
formation model by continuing the study of Extremely Red Objects
(EROs) started with \citet{eros1} (Paper I). EROs are massive, red
galaxies at $0.7\lesssim z\lesssim 3$ and their numbers and properties
have posed a challenge to hierarchical galaxy formation models. In this paper we have analysed the halo
occupation distribution (HOD) and clustering of EROs predicted based on the published model of galaxy formation of
\citet{bower06}. The parameters in this model were set to reproduce
observations of the local galaxy population. No parameter has been
changed for the work presented
here. The \citeauthor{bower06} model gives an impressively close match to the number
counts of EROs \citepalias{eros1}. This model also matches the evolution of the K-band
luminosity function and the inferred evolution of the stellar mass
function. 

The HOD predicted by the \citeauthor{bower06} model has a strikingly
different form from the canonical assumption of a step function that
reaches unity for the central galaxies plus a power
law for satellites.
The central galaxy HOD has a somewhat more rounded form than a step
function and does not
reach unity for K-selected galaxies or EROs. This is due to
the non-monotonic relation between host halo mass and central galaxies
luminosity. According to the \citeauthor{bower06} model, at $z=2.1$, galaxies with $K \leq
20$ are very bright, $L \gtrsim L*$. For such central galaxies, the AGN
feedback in this model truncates the initially monotonic
mass-luminosity relation and increases the scatter, modifying the
shape of their HOD. 
Our predictions suggest that a revision is needed to the canonical
form assumed for the HOD to model the clustering of bright galaxies.

The HOD of EROs suggests that the $(R-K)$ or $(i-K)$ colour cuts do not select all
galaxies above a certain mass threshold since they leave out some of
the younger star forming galaxies. However, EROs with $(R-K)>5$ or
$(i-K)>4.5$ come very close to representing the whole population of
galaxies with $K\leq 20$ at $z\geq 1.5$. On average, halos more massive than $10^{13}
h^{-1}M_{\odot}$ host at least one ERO with $K\leq 20$ and
$(R-K)>5$.

At a given redshift, the contribution of satellites dominates the
shape of the HOD of the redder
EROs. The satellite contribution also becomes increasingly dominant at lower redshifts. 
We predict that, in the \citeauthor{bower06} model, brighter
and redder EROs should be found in more
massive haloes. 

We have found that the minimum halo mass needed to host two EROs can be more than an order of magnitude larger than that needed to host
just one ERO. A similar factor is found for K-selected galaxies at
$1\le z \le 2.5$, so this result is not peculiar to EROs but rather is
intrinsic to bright galaxies.

We predict from the \citeauthor{bower06} model that the spatial two-point correlation function
of EROs is roughly consistent with a power law for scales
$0.03\lesssim r\, \lesssim 30\, h^{-1}{\rm Mpc}$, though in
real-space there is a $\sim 10$\% change of slope  at $r\sim 1\,
h^{-1}{\rm Mpc}$, which reflects the shift in dominance from the
one-halo to the two-halo term. EROs with $(R-K)>5$ and $K<20.2$ at $z=2.1$ are predicted to have a bias of
$\sim 3$.

The predicted clustering does not depend strongly on apparent
magnitude over the range $18.4\leq K \leq 20.2$. We predict the
clustering of redder EROs to be boosted, particularly on small
scales, due to a higher proportion of satellite galaxies. We
have found that those EROs predicted to be experiencing a star formation
burst are more clustered than those which are passively
evolving. However, when separating EROs by their specific star
formation rate, $SFR/M=10^{-11}{\rm yr}^{-1}$, active EROs are predicted to
have a correlation length of $r_{0}=5.7h^{-1}{\rm Mpc}$, which is lower than
that predicted for quiescent ones, $r_{0}=7.7h^{-1}{\rm Mpc}$. The latter value
is comparable but lower than the observational estimations. 

The predicted spatial correlation length of EROs is smaller than most
observational estimates. However, since the match between the predicted angular clustering and
the observed one for EROs with $(R-K)>5$ is rather good, it is quite
likely that the discrepancy in the correlation length is due to the
assumptions required to derive this quantity from the observations. Most of the
observational studies of EROs rely on photometric redshifts, which
introduce a large uncertainty into the redshift distribution, which is
needed to obtain the spatial correlation function from the angular one. The surveyed area usually is
quite small and the integral constraint has to be included to account
for the bias in the inferred mean galaxy density and the impact this
has on the estimated clustering. Another simplification that is usually made is to assume that the
spatial correlation function is a pure power law. Our model predicts departures from a power law, in agreement with observations of
galaxies at low redshifts. Often it is also assumed that the
clustering varies monotonically with redshift, and this evolution is usually parametrised as a
global, scale independent factor. However, our results suggest that the
evolution of $\xi$ with redshift is not monotonic and furthermore that it varies
with pair separation. 

The predicted angular correlation function for
EROs with $(R-K)>5$ matches the observational estimates within the
range of error allowed by sampling variance. On a scale of
$\theta\sim0.02$ deg, we find a small change of slope in the angular clustering due to the transition from
the one-halo to the two-halo terms.

We have also explored the predicted angular clustering of EROs
selected by their $(i-K)$ colour. This colour selection is more
efficient at leaving out galaxies with $z<1$. We find that, unlike the
case of EROs with $(R-K)>5$, the angular clustering of EROs with $(i-K)>4.5$
is underpredicted, particularly at large separations. We have shown that
this difference is due partly to sampling variance. However, at large
scales sampling variance alone cannot explain the factor of 3
difference between the model and the observations. We have explored the possible origin of this discrepancy. We have shown
that the impact of using the redshift distribution estimated directly from
observations, using different i-band filters or doubling the default stellar
yield in the \citeauthor{bower06} model has a minimal impact on the
predicted angular correlation at large scales. We have also run the
\citet{font08} model, finding that, on large scales, it predicts an angular correlation function only
slightly higher in amplitude than that from the \citeauthor{bower06}
model. Unlike in the observations, EROs with $(i-K)>4.5$ are predicted to be dominated
by ``old" galaxies, classified as such by their location in the
colour-colour space proposed by \citet{fang09}. However both the predicted
and the observed clustering for ``old" and ``dusty" EROs are very
similar, and thus a change in the split between these two types cannot
account for the discrepancy in the angular correlation function on
large scales between model and observations. Nevertheless, the problem appears to be related to the
i-band filter, since our prediction for EROs with $(i-K)>4.5$
perfectly matches the observed angular clustering of EROs with $(i-K)>4$.

Overall, the predictions based on the \citeauthor{bower06} model match
observations reasonably well, once sample variance in the current data
is taken into account. 

This is the second paper in a series which examines the properties and
nature of red galaxies in hierarchical models. In the
third paper we will examine different colour cuts used to select red
galaxies at $z>1$ and compare the properties of their present-day descendants.

\subsection*{ACKNOWLEDGEMENTS}

We thank R. Angulo for providing the dark matter correlation function predicted
using the Millennium Simulation. Thanks to N. Roche and X. Kong for
their tabulated two-point angular correlation function. We also thank
A. Edge, J. Helly and F. J. Castander for helpful comments and G. Altay for his
computational tips. The calculations for this paper were performed on the ICC Cosmology Machine, which is part of
the DiRAC Facility jointly funded by STFC, the Large Facilities
Capital Fund of BIS, and Durham University. VGP is supported by a Science and Technology Facilities 
Council rolling grant and acknowledges past support from the Spanish
Ministerio de Ciencia y Tecnolog\'{i}a.

\vspace{-0.7cm}

\bibliographystyle{mn2e}


\end{document}